\documentstyle[twoside,fleqn,espcrc2,epsfig]{article}

\def\cD{{\cal D}}

 \def\a{\alpha} 
\def\b{\beta}
\def\d{\delta}          
\def\De{\Delta}
\def\ga{\gamma}         
\def\gm{\Gamma}
 
\def\la{\lambda}        
\def\La{\Lambda}
\def\ka{\kappa}
\def\m{\mu}
\def\n{\nu}
\def\r{\rho}
\def\s{\sigma}

\def\th{\theta}
\def\eps{\epsilon}
\def\ee{\varepsilon}
\def\om{\omega}

\def\oom{\bar\omega}

 \def\parslash{{\partial{\hspace{-6pt}}/\hspace{1pt}}}
\def\Aslash{{A\hspace{-7pt}/\hspace{2pt}}}
\def\Vslash{{V\hspace{-9pt}/}\hspace{5pt}}
\def\Dslash{{D\hspace{-7pt}/\hspace{2pt}}}

\def\qslash{{q\hspace{-5pt}/}}
\def\kslash{{k\hspace{-5.5pt}/\hspace{0.5pt}}}
 
\def\hg{\hat{g}}
\def\tp{\tilde{p}}
\def\hp{\hat{p}}
 
\def\YM{\rm YM}
\def\CS{\rm CS}
\def\GF{\rm GF}
\def\ES{\rm ES}
 
\def\idx{\int\! d^3\!x \,}
\def\idxth{\int\! d^3\!x \,d^2\!\th \,}
 
\def\igual{\hspace{-7pt}=\hspace{-7pt}}
\def\mas{\hspace{-7pt}+\hspace{-7pt}}
\def\menos{\hspace{-7pt}-\hspace{-7pt}}
\def\ds{\displaystyle}
\def\ss{\scriptstyle}
\def\to{\rightarrow}
\def\igual{\hspace{-7pt}=\hspace{-7pt}}
\def\mas{\hspace{-7pt}+\hspace{-7pt}}
\def\menos{\hspace{-7pt}-\hspace{-7pt}}

\def\unit{{\rm 1\hspace{-2pt}l}}

\newcommand{\AmS}{{\protect\the\textfont2
  A\kern-.1667em\lower.5ex\hbox{M}\kern-.125emS}}

\title{
\vspace{-15mm}
{\normalsize\null\hfill HD-THEP-97-01\\
\vspace{-4mm}
\null\hfill ITP-SB-97-06\\
\vspace{-4mm}
\null\hfill hep-th/9701052\\
}
       Supersymmetric Yang-Mills-Chern-Simons theory\thanks{To
       appear in the Proceedings of the 30th Ahrenshoop Symposium
       on the Theory of Elementary Particles, Nuclear Physics B,
       Proceedings Supplement, edited by D. Lust, 
       H.-J. Otto and G. Weigt.}}

\author{F. Ruiz Ruiz\address{Institut f\"ur Theoretische Physik,
          Universit\"at Heidelberg\\ Philosophenweg 16, 
          69120 Heidelberg, Germany}%
          \thanks{%f.ruiz@thphys.uni-heidelberg.de. 
                  Alexander von Humboldt Fellow.}
        and
        P. van Nieuwenhuizen\address{Institute for Theoretical
          Physics, State University of New York at Stony Brook\\ 
          Stony Brook, NY 11794-3840, USA}%
          \thanks{%vannieu@insti.physics.sunysb.edu. 
                  Research supported by NSF grant Phy 9309888.}}
       
\begin{document}
\thispagestyle{empty}
\begin{abstract}
  We prove that three-dimensional $N\!=\!1$ supersymmetric
  Yang-Mills-Chern-Simons theory is finite to all loops. This leaves
  open the possibility that different regularization methods give
  different finite effective actions. We show that for this model
  dimensional regularization and regularization by dimensional
  reduction yield the same effective action.
\end{abstract}
\maketitle

\section{Introduction and conclusions} 

The supersymmetric regularization of gauge theories remains a major
unsolved problem in supersymmetry. In this contribution we consider
supersymmetric Yang-Mills-Chern-Simons theory in three dimensions and
prove that ordinary dimensional regularization (or DReG) \cite{tHooft}
and regularization by dimensional reduction (or DReD)
\cite{Siegel-DReD} preserve both supersymmetry and BRS invariance. We
further show that they give the same Green functions. Our strategy is
to first prove that the theory is finite to all loop orders, so that
the regularized effective actions $\gm^{\rm DReG}$ and $\gm^{\rm
  DReD}$ are also renormalized effective actions and the difference
$\De\gm=\gm^{\rm DReG} - \gm^{\rm DReD}$ is the difference of two
renormalized effective actions. Next we show that this difference
vanishes. This, together with the observations that DReG preserves at
all stages the BRS identities of local gauge invariance and that DReD
preserves supersymmetry, implies the thesis.  We will also see that
neither DReG nor DReD have problems in dealing with the
$\eps^{\m\n\r}$ that appears in the classical action of the model. The
work reported here is based on ref. \cite{RvN}, to which we refer for
more details.

\section{Superfields and components}

In three-dimensional $N\!=\!1$ supersymmetry, superspace is
parameterized by three real spacetime coordinates $x^\m$ and two real
anticommuting Majorana spinor coordinates $\th^\a$. Any vector $v^\m$
can be represented as a symmetric rank-two vector $v^{\a\b}$ with
indices $\a,\b=1,2$ through the relation $v^\a_{~\b} = ( \ga_\m
)^\a_{~\b}\, v^\m$, where $\ga_\m$ are the Dirac matrices. The spinor
superderivative $D_\a$ is defined by $D_\a = \partial_\a\!  +
i\,\th^\b\partial_{\b\a}.$ Supersymmetry transformations are generated
by the supercharge $Q_\a =\partial_\a\!  - i\,\th^\b\partial_{\b\a}$
and have the form $\d x^{\a\b}\!=\!a^{\a\b}\!\!- 2i\, \ee^{(\a}
\th^{\b)}$, $\d\th^a=\ee^\a$, where $a^{\a\b}$ is a real commuting
constant vector, $\eps^\a$ is an anticommuting constant Majorana
spinor and $\ee^{(\a} \th^{\b)}= {1\over 2}\,(\ee^\a \th^\b +
\ee^\b\th^\a).$ A superfield $\Psi(x,\th)$ transforms linearly under
the action of the supercharge: $\d \Psi = \eps^\a Q_\a\Psi$.

The gauge field $A^a_{\a\b}$ of a real, compact, semi-simple Lie
algebra with completely antisymmetric structure constants $f^{abc}$ is
part of a supermultiplet described by a Majorana spinor gauge
potential $\gm^a_\a$ \cite{1001}. The superfield $\gm^a_\a$ defines a
real vector gauge potential $\gm^a_{\a\b}$ and an imaginary spinor
field strength $W^a_\a$ through the equations
\begin{displaymath}
\begin{array}{l} 
  {\ds \gm^a_{\a\b} = D_{(\a} \gm^a_{\b)} 
     + {i\over 2}\> f^{abc} \,\gm^b_\a \gm^c_\b }\\[9pt] 
  W^a_\a = D^\b D_\a \gm^a_\b 
              + i f^{abc} \, \gm^{b\b} D_\b\gm^c_\a \\[6pt]
  \phantom{W^a_\a ~} {\ds  
  - {1\over 3}\,\, f^{abc} f^{cde}\,\gm^{b\b}\,\gm^d_\b\, \gm^e_\a ~.}
\end{array}
\end{displaymath}
Besides $A^a_{\a\b},$ the supermultiplet contains a real scalar field
$H^a$ and two anticommuting Majorana spinors $\chi^a_\a$ and
$\la^a_\a,$ given by the projections onto $\th^\a\!=\!0$
%\newpage
\begin{displaymath} 
\begin{array}{rrl}
  \chi^a_\a = \gm^a_\a \Big\vert &
  {\ds H^a = {1\over 2}\, D^\a \gm^a_\a \Big\vert} & \\[9pt]
  A^a_{\a\b} = \gm^a_{\a\b} \Big\vert & 
  {\ds \la^a_\a = -\, {i\over 2}\, W^a_\a \Big\vert} & \hspace{-5pt} .
\end{array}
\end{displaymath} 

We will work in the supersymmetric Landau gauge, characterized by the
condition $D^\a \gm^a_\a = 0.$ In this gauge, the classical $N\!=\!1$
Yang-Mills-Chern-Simons action has the form
\begin{displaymath}
  \gm_0 = {1\over m}\> S_{\YM} + S_{\CS} + S_{\GF} + S_{\ES} ~,
\end{displaymath}
where
\begin{displaymath}
  S_{\YM} = - {1\over 32\,g^2} \idxth \,W^{a\a} W^a_\a 
\end{displaymath}
\begin{displaymath}
\begin{array}{l}{\ds
  S_{\CS} = {i\over 16\,g^2} \idxth \,\Big[\,
      \big( D^\a \gm^{a\b}\big) \,\big( D_\b \gm^a_\a\big)}\\[9pt]
  \phantom{S_{\CS} `} {\ds +\, {2i\over 3} \,f^{abc}\, 
          \gm^{a\a}\gm^{b\b}\,\big(D_\b\gm^c_\a\big) }\\[9pt]
  \phantom{S_{\CS} `}  {\ds -\, {1\over 6}\, f^{abc} f^{cde} \,\gm^{a\a}\,
          \gm^{b\b}\,\gm^d_\a \, \gm^e_\b  \Big] }
\end{array}
\end{displaymath} 
\begin{displaymath} 
  S_{\GF}  = {i\over 4} \idxth\, s\, (\, \hat{C}^a  D^\a\gm^a_\a\,)
\end{displaymath}
and 
\begin{displaymath}
   S_{\ES} = {i\over 2} \idxth \,\bigg( \, 
        {1\over 2} \, K^{a\a}_\gm\,s\gm^a_\a - K^a_C\,sC^a \bigg) ~.
\end{displaymath}
Here $m$ is a parameter with dimensions of mass, $g$ is a
dimensionless coupling constant, $s$ is the BRS operator, $\hat{C}^a$
and $C^a$ are real anticommuting antighost and ghost superfields, and
$K^{a\a}_\gm$ and $K^a_C$ are commuting external supersources coupled
to the nonlinear BRS transforms $s\gm^a_\a$ and $sC^a.$ The BRS
transformations that leave $S_{\YM},~S_{\CS}$ and $S_{\GF}$ invariant
are given by
\begin{displaymath}
\begin{array}{ll}   
   s\,\gm^a_\a = i\,\big( \nabla_\a C\big)^a  ~~ &s\,B^a =0 \\[6pt]
   s\,\hat{C}^a = B^a &  
        {\ds s\,C^a= -{1\over 2} \,f^{abc}\, C^b\, C^c ~, }
\end{array}
\end{displaymath} 
with $\nabla^{ab}_\a = \d^{ab} D_\a + i f^{acb}\, \gm^c_\a$ the spinor
covariant superderivative, $B^a$ a real Lagrange multiplier superfield
imposing the gauge condition $D^\a \gm^a_\a = 0$, and $s$ satisfying
as usual $s^2=0$.  The components of $B^a$ and $\hat C^a$ are defined
by the projections
\begin{displaymath} 
\begin{array}{rrl}
     b^a = B^a \,\Big\vert 
   & \qquad\qquad\qquad \hat c^a = \hat C^a \,\Big\vert & 
\\[6pt]
     \zeta^a_\a = i D_\a B^a \,\Big\vert 
   & \hat \varphi^a_\a = D_\a \hat C^a   \,\Big\vert &
\\[6pt]
     {\ds h^a = -\, {\ds {i\over 2}}\, D^2 B^a \,\Big\vert } 
   & {\ds \hat \om^a = 
         -\, {i\over 2} \, D^2 \hat C^a \,\Big\vert  }
   & \hspace{-5pt} ,
\end{array}
\end{displaymath} 
those of $C^a$ by replacing hatted antighosts with unhatted
ghosts, and those of $K^{a\a}_\gm$ and $K^a_C$ by 
\begin{displaymath}
\begin{array}{rrl}
     \ka_\a^a = K^a_{\a\,\gm} \,\Big\vert
   & \qquad\qquad \ell^a = K_C^a \,\Big\vert &
\\[6pt]
     {\ds G^a = -\,{i\over 2}\> D^\a K^a_{\a\,\gm} \,\Big\vert }
   & \tau^a_\a = i\,D_\a K_C^a \,\Big\vert &
\\[9pt]
     K^a_{\a\b} = i\,D_{(\a} K^a_\b{}_{)\,\gm} \,\Big\vert
   & {\ds L^a = -\,{i\over 2}\> D^2 K_C^a \,\Big\vert } &
     \hspace{-5pt} .
\\[6pt]
     {\ds \s^a_\a = -\, {i\over 2}\,
         D^\b D_\a K^a_{\b\,\gm} \,\Big\vert}  & &
\end{array}
\end{displaymath}
By construction, $\gm_0$ is invariant under supersymmetry
transformations. 
%Superpower counting for $\gm_0$ gives a finite number of
%superficially divergent 1PI diagrams, hence proving that the theory
%is superrenormalizable. 

To formulate DReG, we work with component fields. The terms $S_{\CS},~
S_{\YM},~ S_{\GF}$ and $S_{\ES}$ in $\gm_0$ are given in terms of
components by
\begin{eqnarray}
   S_{\YM} &\igual & {\ds {1\over g^2} \idx \, \Big[
       - {1\over 4}\, F^a_{\m\n} F^{a\m\n}
       - {1\over 2} \, \bar\la^a (\Dslash{} \la)^a\,\Big]} 
\nonumber\\[6pt]
   S_{\CS}  &\igual & {\ds {1\over g^2} \idx \,\Big[\,\eps^{\m\n\r}\,
       \Big( \,{1\over 2}\,A^a_\m\partial_\n A^a_\r} 
\nonumber\\[3pt]
   &\,\mas & {\ds {1\over 6}\,f^{abc}\, A^a_\m A^b_\n A^c_\r\, \Big)
       - {1\over 2} \> \bar\la^a \la^a\, \Big] } 
\nonumber\\[6pt]
   S_{\GF}  &\igual & {\ds \idx \> s\, \Big(\! 
       - \hat{c}^a\,\partial_\m V^{a\m}
       + i\, \bar{\hat{\varphi}}^a \La^a - \hat{\om}^a H^a \Big)\, , }
\nonumber
\end{eqnarray}
where $F^a_{\m\n}$ is the field strength,
$D^{ab}_\m=\d^{ab}\partial_\m + f^{acb} A^c_\m$ is the covariant
derivative, and
\begin{eqnarray} 
   V^a_\m &\igual & A^a_\m
      + {1\over 4}\, f^{abc}\,\bar\chi^b\ga_\m \chi^c  
\label{V}\\[6pt]
   \La^a &\igual & \la^a + \parslash\chi^a
      + {1\over 2}\, f^{abc} \,\Aslash^b\chi^c
      - {1\over 2}\, f^{abc}\,H^b \chi^c 
\nonumber \\
   & \menos & {1\over 24}\, f^{abc}\,f^{cde}\, 
      \ga^\m\chi^b \, (\bar\chi^d\ga_\m\chi^e) ~. \label{La} 
\end{eqnarray} 
The action of $s$ on components is obtained from the definition of the
latter as projections and the action of $s$ on superfields. It is
given by
\begin{displaymath}
\begin{array}{ll}
    \gm^a_\a : & s \chi^a =  i\varphi^a - f^{abc} \chi^b c^c \\[6pt]
    & s A^a_\m = (D_\m c)^a \\
    & {\ds s H^a = \om^a + f^{abc} H^b c^c 
                  - {i\over 2}\, f^{abc} \bar{\chi}^b\varphi^c} \\[6pt]
    & s \la^a = - f^{abc} \la^b c^c
\end{array}
\end{displaymath}
{}
\begin{displaymath}
\begin{array}{llll}
    B^a: & sb^a =0 & s\zeta^a=0 & sh^a=0 \\[9pt]
    \hat{C}^a: &  s \hat{c}^a = b^a &  s\hat{\varphi}^a = i\zeta 
               &  s\hat{\om}^a = h^a 
\end{array}
\end{displaymath}
{}
\begin{displaymath}
\begin{array}{ll}
   C^A: & {\ds s c^a = - {1\over 2}\, f^{abc} c^b c^c } \\[6pt]
        & s\varphi^a = f^{abc} \varphi^b c^c \\
        & {\ds s \om^a = - f^{abc} \om^b c^c 
             - {1\over 2}\> f^{abc}\, \bar{\varphi}^b\varphi^c~. }
\end{array}
\end{displaymath}
The supersymmetry transformation laws for the components are obtained
in the same way. Here we only present those for the components of the
gauge multiplet:
\begin{displaymath}
\begin{array}{l}
    \d \chi^a = \Vslash ^a \ee - H^a\ee \\[6pt]
    \d A^a_\m = \bar\ee \ga_\m \la^a + \bar\ee \,(D_\m \chi)^a \\[6pt]
    \d H^a = -\, \bar\ee \La^a \\
    {\ds \d \la^a = -\, {1\over 2}\> \ga^\m\ga^\n F^a_{\m\n}\, \ee
                     + f^{abc} \la^b (\bar\chi^c\ee)~. }
\end{array}
\end{displaymath}

Power counting shows that there is only a finite number of
superficially divergent diagrams, thus proving that the theory is
superrenormalizable. At one loop there are quadratic, linear and
logarithmic divergences; at two loops there are linear and logarithmic
divergences; and at three loops only logarithmic divergences survive.
Furthermore, quadratically divergent one-loop diagrams do not have
internal gauge lines and the only primitively divergent two and
three-loop 1PI diagrams are those in Table 1, where $\oom$ denotes the
superficial UV degree of divergence of the diagram.
\begin{table}[ht]
\vspace{-20pt}
\begin{center}
\begin{tabular}{|l l l l |c|c|} \hline
  \multicolumn{4}{|c|}{external lines}
                     & 2 loops & 3 loops \\ \hline
  $\chi\bar\chi$ &&& &  $\oom=1$ & $\oom=0$  \\ \hline
  $\la\bar\chi$  & $A^2$ & $AH$ & $H^2$ & 
       ${\ds{\atop \oom=0}}$ & \vspace{-4pt} \\
  $\chi\bar\chi A$ & $\chi\bar\chi H$ & $(\chi\bar\chi)^2$ &&&
       \\ \hline
\end{tabular}
\\[9pt]
{\sl Table 1: Power counting for component fields}
\end{center}
\vspace{-20pt}
\end{table}

The BRS identity for the full renormalized effective action $\gm$
takes the form
\begin{displaymath}
\begin{array}{l}
   {\ds \idx \,\bigg(\, 
        \sum_\phi \,{\d\gm\over\d\phi}\,{\d\gm\over\d K_\phi} }\\[3pt]
\phantom{ \ds \idx ~}{\ds
        +\, b\,{\d\gm\over\d\hat{c}}
        + i\bar\zeta\,{\d\gm\over\d\bar{\hat{\varphi}}}
        + h\,{\d\gm\over\d\hat\om}\,\bigg) = 0 ~.}
\end{array}
\end{displaymath}
where the sum is extended over $\phi^a = \chi^a,\, V^a_\m,\,H^a,$ {}
$\La^a,\, c^a,\, \varphi^a,\, \om^a.$ In what follows, we will write
this equation as $(\gm,\gm)=0$ 
% Its solutions\footnote{Since DReG preserves BRS invariance, there is
% at least one solution} are functionals $\gm[\psi,K_\phi]$ of the
% fields $\psi^a = \phi^a, \,\hat{c}^a, \,\hat{\varphi}, \,\hat{\om}$
% and the sources $K^a_\phi$. 
and use the notation $\Theta$ for the Slavnov-Taylor operator:
$\Theta=(\gm_0, ~\,)$. An important property of $\Theta$ is that it
commutes with the supersymmetry generator $\d\!:~[\Theta,\d]=0.$

We remark that $\gm$ generates 1PI Green functions for
the fields $V^a_\m$ and $\La^a$ and not for the elementary fields
$A^a_\m$ and $\la^a.$ This is due to the fact that $S_{\ES}$
introduces external sources for the BRS variations of $V^a_\m$ and
$\La^a,$ and not for those of $A^a_\m$ and $\la^a.$ To compute $\gm$,
we use the Feynman rules for $A^a_\m$ and $\la^a$ and treat $V^a_\m$
and $\La$ as composite fields defined by eqs.  (\ref{V}) and
(\ref{La}). It is not difficult to see that, given a 1PI diagram with
superficial degree of divergence $\oom$, all the diagrams that result
from replacing one or more of the external $A^a_\m$ and/or
$\la^a$-lines with any of the composite fields on the right-hand side
in eqs. (\ref{V}) and (\ref{La}) have superficial degree of divergence
strictly less than $\oom$. Regarding then $V^a_\m$ and $\La^a$ as
composite fields does not worsen power counting.

\section{Dimensional reduction and dimensional regularization}

In DReD, all the fields and matrices are kept three-dimensional and
the momenta are continued in the sense of ordinary DReG to $d\!<\!3.$
Because the Dirac algebra is performed in three dimensions, the Fierz
identities remain valid and DReD manifestly preserves supersymmetry.
The regularized action computed with DReD satisfies then $\d \gm^{\rm
DReD}=0.$ The BRS transformation for the gauge field in DReD, however,
is not the same as in the unregularized theory. Indeed, whereas the
first $d\!<\!3$ components of the gauge field have the same BRS
transformation law as the gauge field in the unregularized theory, the
last $3\!-\!d$ components transform as $s A^a_\m = f^{abc}A^b_\m c^c.$
Due to this fact, DReD does not manifestly preserve BRS invariance. It
may happen that at the end of all calculations, once the limit $d\to
3$ has been taken, all effects due to the splitting of the gauge field
into $d$ and $3\!-\!d$ components go away, but this is not what is
meant by manifest BRS invariance.
%As a matter of fact, here we are going to show that indeed DReD's
%splitting does not modify the BRS identities when the limit $d\to 3$
%is taken. 
Concerning the well known algebraic inconsistency \cite{Siegel-incon}
that occurs in products of three or more epsilons in DReD, we mention
that it disappears in the limit $d\to 3,$ since contributions with
three or more epsilons are finite by power counting at $d\!=\!3.$

To define DReG, we follow ref. \cite{GMR} and treat $\eps^{\m\n\r}$ as
purely three-dimensional \cite{tHooft}. This gives for the propagator
of the gauge field in $d\!\geq\! 3$ dimensions
\begin{equation}
  \De_{\m\n}(p) = D_{\m\n}(p) + R_{\m\n}(p) ~,
\label{prop}
\end{equation}
where
\begin{displaymath}
  D_{\m\n}(p) = - \, g^2 m ~ 
    {m\, \eps_{\m\r\n}\, p^\r + i\,p^2 g_{\m\n} - i\, p_\m p_\n 
     \over p^2 \,(p^2\!+\!m^2\!-\!io)} 
\end{displaymath}
\begin{displaymath}
\begin{array}{l} {\ds
   R_{\m\n}(p) = -\,
     {g^2 \,m^3 \over (p^2\!-\!io)^2 + m^2\,\tp^2}\,
       \bigg[ \, {1\over p^2\!+\!m^2\!-\!io} }\\[12pt]
\phantom{~~} 
  \times \, {\ds {\hp^2\over p^2}\, 
        \Big( m\,\eps_{\m\r\n}\,p^\r + i\, p^2 g_{\m\n}
        + {i\,m^2\over p^2\!-\!io} \,p_\m p_\n\Big) } \\[3pt]
\phantom{~~\,} 
   {\ds +\, {i \over p^2\!-\!io} ~ \Big(\, \tp^2 \hg_{\m\n}
          - p_\m \hp_\n - \hp_\m p_\n  + \hp_\m \hp_\n \Big)\,
          \bigg] ~  }
\end{array}
\end{displaymath}
Here $g_{\m\n}$ and $p^\m$ are $d$-dimensional, objects with a tilde
are three-dimensional and objects with a caret are
$(d\!-\!3)$-dimensional. Since the propagator is the inverse of the
kinetic term in the $d$-dimensional classical action and the BRS
transformation for the gauge field is the same as in the regularized
theory, DReG preserves BRS invariance \cite{GMR} \cite{Breitenlohner}.
Hence, the DReG regularized effective action satisfies the BRS
identity
%\begin{displaymath}
$(\gm^{\rm DReG},\gm^{\rm DReG})=0.$
%\end{displaymath}
The complicated propagator for the gauge field is the price for having
a consistent treatment of $\eps^{\m\n\r}$ while manifestly preserving
BRS invariance. As regards supersymmetry, it is well known that DReG
does not manifestly preserve it.

\section{Perturbative finiteness}

To prove perturbative finiteness at one loop, we consider a one-loop
1PI diagram and denote by $\cD(d)$ its value in DReG. According to eq.
(\ref{prop}), if the diagram has an internal gauge line, $\cD(d)$ is
the sum of two contributions: $\cD(d) = \cD_D(d) + \cD_R(d)$. The
contribution $\cD_D(d)$ contains the $SO(d)$ covariant part
$D_{\m\n}$ of all the gauge propagators. The contribution $\cD_R(d)$
contains at least one $R_{\m\n}$ and can be easily seen to be both UV
and IR finite at $d\!=\!3$ by power counting. Being finite at $d\!=\!
3$ and being at least linear in $\hg_{\m\n},$ $\cD_R(d)$ vanishes as
$d\to 3$. We are thus left with only the $SO(d)$ covariant $\cD_D(d)$.
If the diagram has no internal gauge line, $\cD(d)$ is already $SO(d)$
covariant. Using that $SO(d)$ covariant one-loop integrals have no
poles when $d$ is continued to a positive odd integer \cite{Speer}
completes the proof at one loop. This also proves that in the limit
$d\to 3,$ 1PI Green functions at one loop are identical in DReG and in
DReD.

At two loops we proceed differently. Let us assume that the two-loop
correction $\gm^{\rm DReG}_2$ to the effective action consists in the
limit $d\to 3$ of a divergent part $\gm^{\rm DReG}_{\rm 2,div}$ and a
finite part $\gm^{\rm DReG}_{\rm 2,fin}.$ Since $\gm^{\rm DReG}_2$
satisfies the BRS identity
\begin{equation}
  \Theta \gm^{\rm DReG}_2 
     + (\gm^{\rm DReG}_1,\gm^{\rm DReG}_1) = 0 
\label{BRS-id-2}
\end{equation} 
and $\gm^{\rm DReG}_1$ is finite, the divergent part $\gm^{\rm
  DReG}_{\rm 2,div}$ satisfies $\Theta \gm^{\rm DReG}_{\rm 2,div}=0.$
Because 1PI Feynman diagrams with external sources as external lines
are finite by power counting and there are no one-loop subdivergences,
$\gm^{\rm DReG}_{\rm 2,div}$ does not depend on the external sources
and $\Theta \gm^{\rm DReG}_{\rm 2, div}=0$ reduces to $s\gm^{\rm
  DReG}_{\rm 2, div} =0.$ Using the power counting in Table 1 and that
contributions to two-loop 1PI diagrams from $R_{\m\n}$ are finite, we
have that the most general form of $\gm^{\rm DReG}_{\rm 2,div}$ is
$\gm^{\rm DReG}_{\rm 2,div} = {1\over d-3} P_{\oom_2}$, where 
\begin{equation}
\begin{array}{l}
   P_{\oom_2} =  m {\ds \idx \,\Big[\, 
       \a_1\,m\, \bar{\chi}^a \chi^a 
     + \a_2 \,\bar{\chi}^a \parslash \chi^a }
\\[3pt]\phantom{P_{\oom_2}\,} 
   + \,\a_3 \,\bar{\chi^a} \la^a + \a_4\, A^a A^a 
\\[3pt]\phantom{P_{\oom_2}\,} 
   + \, \a_5\, H^a H^a  
     + \a_6\, f^{abc} \bar{\chi}^a\Aslash^b\chi^c
\\[3pt]\phantom{P_{\oom_2}\,}
   +\, \a_7\, f^{abc} f^{cde} (\bar\chi^a\ga^\m\chi^b)\,
                  (\bar\chi^d\ga_\m\chi^e)\,\Big]  
\end{array}
\label{P2}
\end{equation} 
%\newpage
and $\a_1,\ldots,\a_7$ are numerical coefficients. 
%The terms in $P_{\oom_2}$ correspond to all two-loop Lorentz
%invariant divergences that can be constructed from Table 1 with
%$\oom_2$ derivatives. 
The equation $s\gm^{\rm DReG}_{\rm 2,div}=0$ is an equation in the
coefficients $\a_i$ whose only solution is $\a_i=0$. This completes
the proof at two loops.

The proof at three loops is analogous. Now the only three-loop Lorentz
invariant divergence is $\gm^{\rm DReG}_{3,div}={1\over d-3}
P_{\oom_3}$, with
\begin{displaymath}
 P_{\oom_3} = \a m^2 \idx \bar\chi^a\chi^a ~,
\end{displaymath}
but $P_{\oom_3}$ is not BRS invariant. At higher loops, finiteness
follows from power counting and from absence of subdivergences.

\section{A BRS invariant and supersymmetric effective action}

Since the theory is finite, every regularization method defines a
renormalization scheme. We consider two renormalization schemes:
scheme one uses DReG as regulator and performs no subtractions, scheme
two uses DReD and performs no subtractions. We want to prove that the
difference $\De\gm= \gm^{\rm DReG} - \gm^{\rm DReD}$ between the
corresponding renormalized effective actions is zero. We have seen in
section 4 that this is the case at one loop. So let us consider the
two-loop case.

There is a general theorem in quantum field theory \cite{Hepp} that
states that if two different renormalization (not regularization)
schemes yield the same Green functions up to $k\!-\!1$ loops, then at
$k$ loops they give Green functions that can differ at most by a local
finite polynomial in the external momenta of degree equal to the
superficial overall UV degree of divergence $\oom_k$ at $k$ loops.
This, and the power counting in Table 1, implies that the difference
$\De\gm_2$ at two loops can at most be of the form
\begin{equation}
 \gm^{\rm DReG}_2 - \gm^{\rm DReD}_2 = P_{\oom_2} ~,
\label{diff-2}
\end{equation} 
with $P_{\oom_2}$ as in eq. (\ref{P2}). 
% Since DReG preserves BRS invariance, $\gm^{\rm DReG}_2$ satisfies
% the BRS identity (\ref{BRS-id-2}) and, 
We observe that, since DReD preserves supersymmetry, 
$\gm^{\rm   DReD}_2$ satisfies
\begin{equation} 
  \d \gm^{\rm DReD}_2 = 0 ~ .
\label{susy-id-2}
\end{equation}
Acting with $\d$ on eq. (\ref{BRS-id-2}), using eqs. (\ref{diff-2})
and (\ref{susy-id-2}), and recalling that $[\Theta,\d]=0$ and that
$\De\gm_1=0$, we obtain that $\Theta \d P_{\oom_2}=0$. Since
$P_{\oom_2}$ does not depend on the external sources, $\d P_{\oom_2}$
is independent of the external sources and $\Theta\d P_{\oom_2}=0$
reduces to $s\d P_{\oom_2}=0,$ which is an equation in the
coefficients $\a_i$ in $P_{\oom_2}$. Because $\d P_{\oom_2}$ depends
polynomially on the components of the gauge multiplet and their
derivatives and has an overall factor of $m,$ any nontrivial $\d
P_{\oom_2}$ satisfying $s\d P_{\oom_2}=0$ must be $m$ times a local
BRS invariant of mass dimension two.  However, there are no such
invariants. Hence, $\d P_{\oom_2}=0.$ The only supersymmetry invariant
that can be formed from $P_{\oom_2}$ is
\begin{displaymath}
\begin{array}{l}
  {\ds P_{\oom_2}^{\rm susy} = \a\, m \idx \, \Big[\, 
         {1\over 2}\, \bar{\chi}^a \parslash \chi^a }\\[9pt]
\phantom{  P_{\oom_2}^{\rm susy} \,}
       +\, \bar{\chi^a} \la^a  + A^a A^a -H^a H^a  \\[3pt]
\phantom{  P_{\oom_2}^{\rm susy} \>}
  {\ds - \,{1\over 48} \> f^{abc} f^{cde} (\bar\chi^a\ga^\m\chi^b)\,
                  (\bar\chi^d\ga_\m\chi^e)\,\Big] \, .}
\end{array}
\end{displaymath}
At this point we have exhausted all the information given by BRS
symmetry and supersymmetry. We determine the value of the coefficient
$\a$ in $P_{\oom_2}^{\rm susy}$ by means of an explicit calculation
(see below) and find $\a\!=\!0.$

% The only way left to determine the value of the coefficient $\a$ in
% $P_{\oom_2}^{\rm susy}$ is to compute it using Feynman diagrams. We
% do this below and find that $\a\!=\!0.$ 

At three loops, the difference is $\Delta \gm_3 = \a
P_{\oom_3}$. Since $\Delta\gm_3$ is not BRS invariant, nor
supersymmetric, the same arguments as used at the two-loop level are
now powerful enough to conclude that $\a\!=\!0$ without the need of
any explicit computation.  At higher loops, the difference $\De\gm$
vanishes since at one, two and three loops it vanishes and there are
no overall divergences by power counting.

We now compute $\a$ in $P_{\oom_2}^{\rm susy}.$
To do this, we evaluate the difference between the contributions
from DReG and DReD to the selfenergy of the field $H^a.$ The vertices
with an $H$ are $ H\zeta\chi,~ H\hat{\varphi}\varphi,~ H\hat{\om} c$
and $H\hat{\varphi}\chi c.$ Using them, one can construct two-loop 1PI
diagrams with the six topologies in Fig. 1. In fact, since
$\hat{\varphi}$ only propagates in $\varphi$ and $c$ into $\hat{c},$
and there is no four-point vertex containing the fields $H,\,\varphi$
and $\hat{c},$ no graphs with the topology of Fig. 1a can be
constructed. The topologies in Figs. 1b and 1c, being products of
one-loop topologies, give the same contributions in DReG as in DReD,
hence they do not contribute to $\a.$ We are thus left with the
topologies in Figs. 1d, 1e and 1f. Because one-loop subdiagrams give
the same contributions in DReG as in DReD, only the
\begin{figure}[ht]
\vspace{10pt}
\begin{center}
\epsfig{file=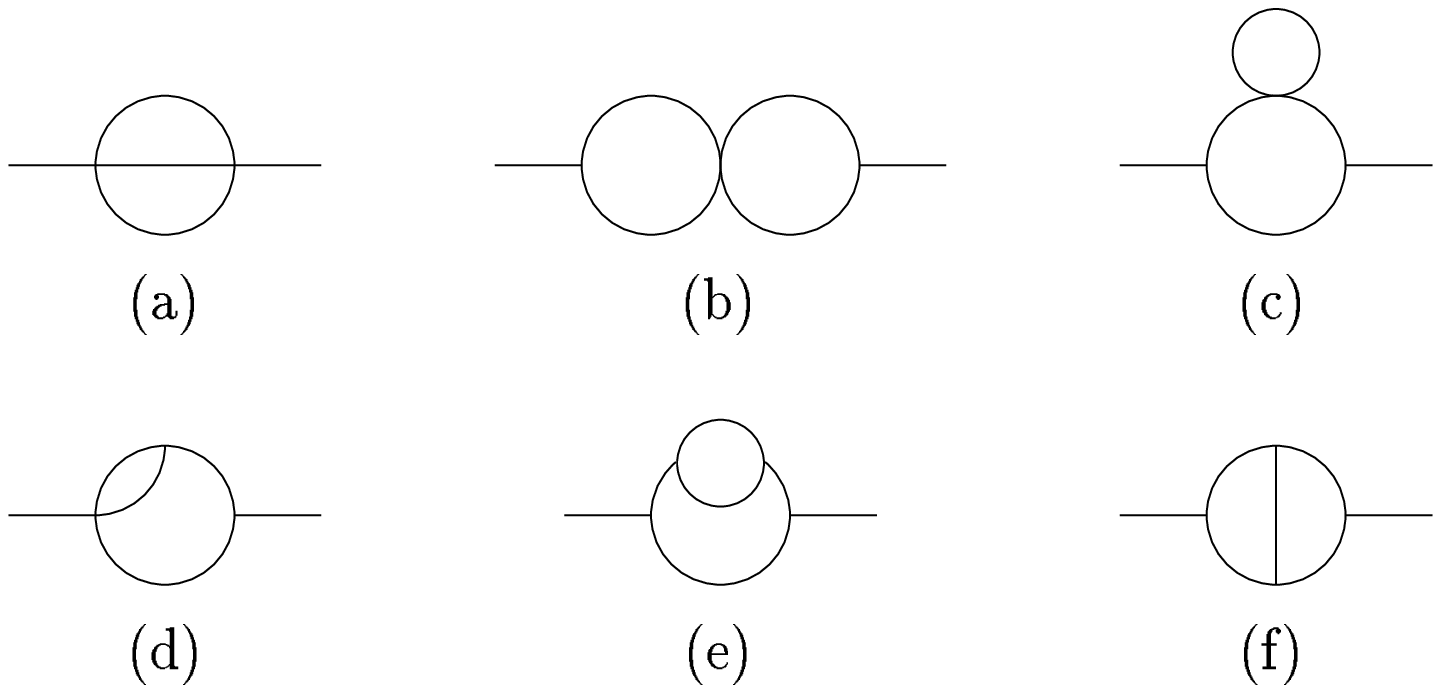,width=.35\textwidth}
\\
{\sl Figure 1: Two-loop topologies for $\langle HH\rangle_{\rm 1PI}$} 
\end{center}
\vspace{-20pt}
\end{figure}
overall divergent part of the corresponding two-loop diagrams
contribute to $\a.$ Since the two-loop diagrams are logarithmically
divergent, the contributions to $\a$ come from setting in the
numerators the external momentum $p^\m$ and the mass $m$ equal to
zero, except, of course, for the overall factor $m.$ The overall
divergent part of every diagram then reads
\begin{displaymath}
   m \int \!{d^dk\over (2\pi)^d} ~ {d^dq\over (2\pi)^d}~ 
      { N(k,q) \over D(k,q,p,m)} ~.
\end{displaymath}
It is very easy to see that the numerator $N(k,q)$ always contains a
trace over a fermion loop. This, and the observation that diagrams
with internal gauge lines only occur in topology 1e and that their
contributions separately cancel, implies that the overall divergence
in DReG and DReD are the same except for the trace over the fermions.
The trace of a sum of products of $\qslash$ and $\kslash$ can always
be written as $d$-dimensional scalar products $k^2,\,kq$ and $q^2$
times an overall trace of the unit matrix. After summing over
diagrams, $\a$ can then be written as
\begin{displaymath}
\begin{array}{l}
   \a = (\, {\rm tr}_{\ss\rm \,DReG} \, \unit 
            - {\rm tr}_{\ss\rm \,DReD} \,\unit\, )\\[6pt]
\phantom{\a~}{\ds \times 
      \int \!{d^dk\over (2\pi)^d} ~ {d^dq\over (2\pi)^d}~ 
      { f(k^2,kq,q^2) \over D(k,q,p,m)} ~,}
\end{array}
\end{displaymath}
where $f(k^2,kq,q^2)$ is a polynomial of its arguments. Because the
theory is finite, the integral is finite and therefore the difference
due to the trace vanishes in the limit $d\to 3.$ Hence $\a=0.$

The equality of $\gm^{\rm DReG}$ and $\gm^{\rm DReD}$ is not explained
by local quantum field theory. One possible explanation might be that
there exists a third, as yet unknown, symmetry of the model.  Another
explanation might be that the existing theorems of local quantum field
theory \cite{Hepp} concerning the difference between the renormalized
expressions for the same Green function computed in two different
renormalization schemes can be sharpened for finite models which are
superrenormalizable by power counting and which have symmetries.

Our analysis relies on the fact that our model is superrenormalizable
by power counting and finite. There exist several all-loop
\cite{Ermushev} finite supersymmetric models in four dimensions, and
$N\!=\!4$ Yang-Mills theory is also all-loop finite. It would be
interesting to apply the methods developed in this paper to these
models (see ref. \cite{CJvN} for a partial comparison of DReG and DReD
in 4-dimensional $N\!=\! 1$ Yang-Mills theory in a non-supersymmetric
gauge).

\end{document}